\documentclass[twocolumn,10pt]{revtex4}

\usepackage{amsmath}
\usepackage{graphicx}
\usepackage{array}
\usepackage{multirow}

\begin{document}

\title{Nonperiodic oscillation of bright solitons in the condensates with a periodically oscillating harmonic potential}
\author{Z. M. HE, D. L. WANG, Y. C. SHE, and J. W. DING$\footnote{%
 E-mail :jwding@xtu.edu.cn}$} \affiliation{$$Department of Physics ${\&}$
Institute for Nanophysics and Rare-earth Luminescence, Xiangtan
University, Xiangtan 411105, China}
\date{\today }

\vskip .2cm
\begin{abstract}

 Considering a
periodically oscillating harmonic potential, we explore the
dynamics properties of bright solitons in a Bose-Einstein
condensate. It is found that under a slower oscillating potential,
soliton movement exhibits a nonperiodic oscillation while it is
hardly affected under a fast oscillating potential. Furthermore,
the head-on and/or ``chase" collisions of two solitons have been
obtained, which can be controlled by the oscillating frequency of
potential.
 \vskip .5cm
{\noindent\emph{Keywords\emph{:}}} Bose-Einstein condesates,
oscillating solitons, periodically oscillating potential trap,
Darboux transformation
\end{abstract}

\maketitle

\vskip 1.5cm
 {\noindent \textbf{1. Introduction}}

\vskip .3cm

Bose-Einstein condensates (BECs) in weakly interacting alkali
atomic gases have proved to an ideal laboratory system for
investigating fundamental nonlinear phenomena, such as bright
solitons,$^{1-3)}$ dark solitons,$^{4-8)}$ and vortices.$^{9)}$
Especially, the bright solitons in BECs open possibilities for
future applications in coherent atomic optics, atom
interferometry, and atom transport.$^{2)}$ Recently, the soliton
oscillations have been obtained in the experiment,$^{10-13)}$
which boost an immense theoretical interest in the nonlinear
matter waves.

Theoretically, it was shown that the external potentials have an
important effect on oscillating properties of solitons in the
one-dimensional BECs.$^{14-20)}$ For example, when a bright
soliton is loaded into an attractive harmonic potential, it
executes harmonic oscillations, of which the oscillating frequency
depends on the trapping frequency.$^{15,19)}$ For an optical
potential, when the energy of bright soliton is lower than the
height of the potential, it can exhibit oscillating behavior
around the bottom of potential notch, of which the oscillating
frequency depends on both the lattice spacing and height of
potential.$^{19)}$ Similar results have also been obtained when
the bright solitons are loaded into a tanh-shaped
potential.$^{19)}$ Usually, the periodic oscillation of bright
solitons can be expected under a spatially nonuniform potential
trap.$^{15,19)}$ In fact, a periodically oscillating external
potential is easily achieved in BEC experiment.$^{21, 22)}$ In
this case, how about the dynamics behaviors of bright solitons? To
our knowledge, there is little report on this subject.

In this paper, we explore the oscillating properties of bright
solitons in BEC with a periodically oscillating harmonic
potential. A nonperiodic oscillating behavior of bright soliton is
obtained under a slow oscillating potential, different from that
under a spatially nonuniform potential trap. Also, the head-on
and/or ``chase" collisions of two solitons have been observed,
which can be controlled by the oscillating frequency of potential.
The results are very useful for future applications of BEC in
accurate atomic clocks and other devices.

\vskip 0.3cm

{\noindent \textbf{2.} \textbf{Model and the Soliton Solutions}}

 \vskip 0.3cm

A periodically oscillating harmonic potential$^{23-25)}$ can be
given by $V(r)=m\omega^{2}_{\perp}
(Y^{2}+Z^{2})+m\omega^{2}_{1}[X-k\sin(\omega_{\perp}\omega
T/2)]^{2}/2$. Here \emph{m} is atomic mass; $\omega_{\perp}$ and
$\omega_{1}$ are the radial and transverse trapping frequencies,
respectively;  \emph{k} and $\omega$ are oscillating amplitude and
oscillating frequency of the external potential, respectively. If
$\omega_{\perp}\gg|\omega_{1}|$, it is reasonable to reduce
Gross-Pitaevskii equation into one-dimensional nonlinear
Schr\"{o}dinger equation with an oscillating harmonic potential
\begin{equation}
i\psi_{t}+\psi_{xx}+2g\mid\psi\mid^{2}\psi-\frac{\omega^{2}_{1}}{\omega^{2}_{\perp}}[x-k\sin(\omega
t)]^{2}\psi=0,
\end{equation}
where $g=-2Na_{s}/a_{\perp}$, and the time t and coordinate x are
measured, respectively, in units of $2/\omega_{\perp}$ and
$a_{\perp}$, with $a_{\perp}=\sqrt{\hbar/m \omega_{\perp}}$. Here,
$N$ is atom number, and $a_{s}$ is \emph{s}-wave scattering length
(SL).$^{26, 27)}$ The tuning of SL can be achieved by Feshbach
resonance.$^{28, 29)}$ We here consider the time-dependent SL
$^{30, 31)}$ $a_{s}=c \exp(\gamma t)$ ( where \emph{c} is constant
and $\gamma^{2}=-4\omega^{2}_{1}/\omega^{2}_{\perp}$).

To obtain the exact solutions of eq. (1), we here make use of the
Darboux transformation.$^{32-34)}$ The seed solution of eq. (1)
can be chosen as
$\psi_{0}=\sqrt{g}Q\exp[ic\exp(2g\gamma)/(2\gamma)]$, where
$Q=\exp(-i\gamma x^{2}/2-iDt-iE)$, with $D=[k\gamma^{3}\sin(\omega
t)+k\gamma^{2}\omega\cos(\omega t)]/(\gamma^{2}+\omega^{2})$ and
$E=\{\sin(2\omega
t)[k^{2}\gamma^{4}\omega^{2}-k^{2}\gamma^{6}-4k^{2}(\gamma^{2}+\omega^{2}]/2+k^{2}\gamma^{5}\omega\sin^{2}(\omega
t)\}/[8\omega(\gamma^{2}+\omega^{2})^{2}]+k^{2}(5\gamma^{2}+4\omega^{2})t
/(8\gamma^{2}+8\omega^{2})$. Subsequently, the Lax-pair of eq. (1)
is presented as
\begin{equation}
\Phi_{x}=U\Phi,      \Phi_{t}=V\Phi,
\end{equation}
where $\Phi=(\phi_{1},   \phi_{2})^{T}$, the superscript ``T"
denotes the matrix transpose. Here,
$U=\left(\begin{array}{cc}\lambda &p\cr -\bar{p}&-\lambda
\cr\end{array}\right)$, and $V=\left(\begin{array}{cc}A&B\cr
C&-A\cr\end{array}\right)$, with $p=\sqrt{g}\psi\bar{Q},
A=2i\lambda^{2}+i\lambda(\gamma x-D)g|\psi|^{2},
B=2i\sqrt{g}\psi\bar{Q}\lambda+i\sqrt{g}\psi_{x}\bar{Q}+\sqrt{g}\psi\bar{Q}(\gamma
x-D)/2$, and
$C=-2i\sqrt{g}\bar{\psi}Q\lambda-i\sqrt{g}\bar{\psi}_{x}Q-\sqrt{g}\bar{\psi}Q(\gamma
x-D)/2$ (the overbar denotes the complex conjugate). From the
compatibility condition $\frac{\partial^{2}\Phi}{\partial x
\partial t}=\frac{\partial^{2}\Phi}{\partial t \partial x}$, one
has $U_{t}-V_{x}+UV-VU=0$. By performing the Darboux
transformation
\begin{equation}
\psi_{1}=\psi_{0}+2(\lambda_{1}+\bar{\lambda}_{1})\frac{Q
\phi_{1}\bar{\phi}_{2}}{\sqrt{g}|\phi_{1}|^{2}+|\phi_{2}|^{2}},
\end{equation}
we can obtain a single soliton solutions of eq. (1)
\begin{equation}
\psi_{1}=[-1+2\frac{(\lambda^{2}_{0}-1)\cos(\varphi)+i\lambda\sqrt{\lambda^{2}_{0}-1}\sin(\varphi)}{\lambda^{2}_{0}\cosh(\theta)+\lambda\sqrt{\lambda^{2}_{0}-1}\sinh(\theta)-\cos(\varphi)}]\psi_{0}.
\end{equation}
Here $\theta=2c\exp(\gamma
t)\sqrt{\lambda^{2}_{0}-1}[x-k\gamma^{2}\sin(\omega
t)/(\gamma^{2}+\omega^{2})+c\exp(\gamma t)/\gamma)]$ and
$\varphi=2c^{2}\exp(2\gamma t)\sqrt{\lambda^{2}_{0}-1}/\gamma$
with $\lambda_{0}$ constant.

Then, by repeating the Darboux transformation \emph{N} times, we
can obtain the \emph{N}-order solution
\begin{equation}
\psi_{n}=\psi_{0}+2\sum^{N}_{n=1}(\lambda_{n}+\bar{\lambda}_{n})\frac{\phi_{1}[n,\lambda_{n}]\bar{\phi}_{2}[n,\lambda_{n}]Q}{\sqrt{g}\Phi[n,\lambda_{n}]^{T}\bar{\Phi}[n,\lambda_{n}]},
\end{equation}
with $\Phi[n,\lambda]=(\lambda I-S[n-1])\cdot\cdot\cdot(\lambda
I-S[1])\Phi[1,\lambda]$. Here
$S_{l_{1}l_{2}}[n']=(\lambda_{n'}+\bar{\lambda}_{n'})\frac{\phi_{l_{1}}[n',
\lambda_{n'}]\bar{\phi}_{l_{2}}[n', \lambda_{n'}]}{|\phi_{1}[n',
\lambda_{n'}]|^{2}+|\phi_{2}[n',
\lambda_{n'}]|^{2}}-\bar{\lambda}_{n'}\delta_{l_{1}l_{2}}, l_{1},
l_{2}=1,2, n'=1,2,\cdot\cdot\cdot,n-1$, and
$n=2,3,\cdot\cdot\cdot, N$. For \emph{N}=2, one can obtain the two
solitons solution of eq. (1)
\begin{equation}
\psi_{2}=\psi_{0}[1+2\frac{G}{F}],
\end{equation}
where
$F=(\lambda_{01}+\lambda_{02})^{2}(h_{1}+k_{1})(h_{2}+k_{2})-4\lambda_{01}\lambda_{02}(h_{1}h_{2}+k_{1}k_{2}+j_{1}j_{2})+2\sqrt{\lambda^{2}_{01}-1}\sqrt{\lambda^{2}_{02}-1}\sin(\varphi_{1})\sin(\varphi_{2})$
and
$G=2\lambda_{02}(\lambda^{2}_{02}-\lambda^{2}_{01})(h_{1}+k_{1})[j_{2}+i
\sqrt{\lambda^{2}_{02}-1}\sin(\varphi_{2})]+2\lambda_{01}(\lambda^{2}_{01}-\lambda^{2}_{02})(h_{2}+k_{2})[j_{1}+i
\sqrt{\lambda^{2}_{01}-1}\sin(\varphi_{1})]$, with
$k_{i}=(2\lambda^{2}-1)\cosh(\theta_{i})+2\lambda_{0i}\sqrt{\lambda^{2}-1}\sinh(\theta_{i})-\cos(\varphi_{i)})$,
$h_{i}=\cosh(\theta_{i})-\cos(\varphi_{i})$, and
$j_{i}=-\lambda_{0i}\cosh(\theta_{i})-\sqrt{\lambda^{2}_{0i}-1}
\sinh(\vartheta_{i})+\lambda_{0i}\cos(\varphi_{i})$ $(i=1,2)$.
Here $\theta_{i}=2c\sqrt(\lambda^{2}_{0i}-1)exp(\gamma
t)[x-k\gamma^{2}\sin(\omega
t)/(\gamma^{2}+\omega^{2})+c\exp(\gamma t)/\gamma)]$ and
$\varphi=2c^{2}\exp(2\gamma t)\sqrt{\lambda^{2}_{0i}-1}/\gamma$
with $\lambda_{0i}$ constant $(i=1,2)$. From eqs.(4) and (6), we
can explore in detail the dynamic behavior of bright solitons.

\vskip 0.3cm

{\noindent \textbf{3. }\textbf{Results and Discussion}}

 \vskip 0.3cm

As a typical example, we consider a BEC consisting of $^{7}$Li.
Based on the currently experimental conditions, the radial and
transverse trapping frequencies are chosen as
$\omega_{\perp}=\pi\times100$ Hz and $\omega_{1}=5\pi i$ Hz,
respectively.$^{2)}$ So, the time and space units correspond to
6.4\emph{ms} and 5.4 $\mu m$  in reality, respectively.

\vskip 0.3cm

{\noindent \emph{3.1}  \emph{Oscillating properties of a single
bright soliton}}

 \vskip 0.3cm
In order to explore the oscillating properties of single soliton
in a $^{7}$Li BEC with a periodically oscillating harmonic
potential, we here propose that the $\omega=10$ and $\omega=0.2$
represent the fast and slow oscillation of harmonic potential,
respectively. Figure 1 shows the space-time distribution of the
density of a BEC under an oscillating harmonic potential.

For a fast oscillating potential, one can see from Fig. 1 (a) that
a bright soliton appears at the initial time. With the time going
on, the amplitude of bright soliton increases but its width
decreases. Meanwhile, the bright soliton propagates along the
positive direction of \emph{x}-axis. This phenomenon is similar to
that of \emph{k}=0 in Ref. [14]. This shows that the propagation
properties of bright solitons are hardly dependent on the fast
oscillating potential.

For a slow oscillating potential, the dynamical properties of the
single soliton are shown in Fig. 1(b). One can see that the bright
soliton moves along the positive direction of \emph{x}-axis when
the time increases from 0 to 10, which is similar to that of the
Fig. 1(a). While the time increases from 10 to 20, interestingly,
it is observed that the bright soliton moves along the negative
direction of \emph{x}-axis, but it can not comeback to the initial
position. When the time further increases, the bright soliton
again moves along the positive direction of \emph{x}-axis. This
indicates that the bright soliton exhibits a nonperiodic
oscillation, different from the periodic oscillation under a
spatially nonuniform potential trap.

Therefore, we conclude from Fig.1 that the noperiodic oscillating
behavior of a bright soliton strong depends on the oscillating
frequency of harmonic potential.
\begin{figure}[tbp]
\includegraphics[width=3.4in]{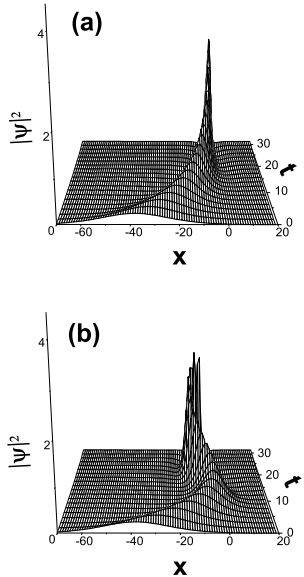}
\caption{The space-time distributions of the density of a BEC with
the harmonic potential showing (a) a faster oscillation
$(\omega=10)$, (b) a slower oscillation $(\omega=0.2)$. The other
parameters used are $\lambda_{0}=2.0$, $c=-0.01$, $\gamma=0.1$,
and $k=50$. }
\end{figure}

\vskip 0.3cm

{\noindent \emph{3.2}  \emph{Oscillating properties of two bright
solitons}}

 \vskip 0.3cm

We further explore in the Fig. 2 the oscillating properties of two
bright solitons under a periodically oscillating potential. We
here choose the $\omega=10$ and $\omega=0.1$ as the typical
examples for the fast and slow oscillating potential,
respectively.

For a fast oscillating potential, one can see from Fig. 2(a) that
there exist two bright solitons at the initial time. With the time
going on, the left bright soliton moves rightward while the right
one moves leftward. Also, the amplitude of each soliton increases
while their width decreases. When the time further increases,
their distance of two solitons further decreases. At $t\approx
20$, the two bright solitons take place a head-on collision. This
phenomenon is similar to that of \emph{k}=0 in Ref. [35].
Therefore, the propagation properties of two bright solitons are
hardly affected by the fast oscillating potential, just as in the
case of a bright soliton strong.

For a slow oscillating potential, the dynamics properties of two
bright solitons are shown in Fig. 2(b). When the time \emph{t}
increases from 0 to 15, two solitons both move along the positive
direction of \emph{x}-axis. While the time t increases from 15 to
20, two solitons both move along the negative direction of
\emph{x}-axis. This shows that two solitons exhibit an oscillating
behavior. Meanwhile, the distance between the two solitons becomes
smaller, two solitons exhibit a ``chase" collision at $t\approx
20$. Obviously, both the head-on and ``chase" collision in Fig. 2
can be controlled by the oscillating frequency of harmonic
potential.

 The validity of the GP equation relies on the condition
that the system be dilute and weakly interacting:
$d|a_{s}(t)|^{3}\ll 1$, where \emph{d} is the  average density of
the condensate. In the real experiment of $^{7}$Li atoms, the
typical value of the atomic densities is 10$^{13}$ cm$^{-3}$. In
our work, we consider that the absolute of the SL is $
|a_{s}(t)|_{max}=30.6 a_{B}$ with $a_{B}$ Bohr radius, so that
$d|a_{s}(t)|^{3}<10^{-4}\ll 1$. Therefore, the GP equation is
valid for the given parameters and thus our results can be
observed under the condition of the current experiments.
\begin{figure}[tbp]
\includegraphics[width=3.4in]{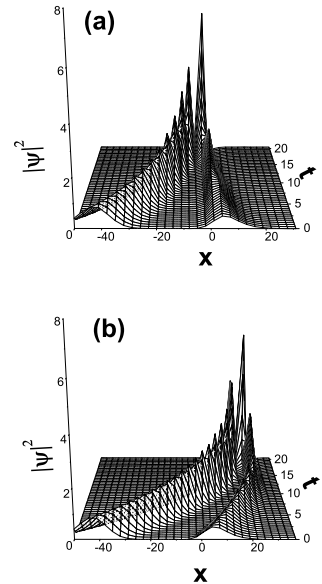}
\caption{The oscillating properties of two bright solitons in a
BEC with the harmonic potential showing (a) a faster oscillation
$(\omega=10)$, (b) a slower oscillation $(\omega=0.1)$. The other
parameters used are $\lambda_{01}=2.0$, $\lambda_{02}=2.5$, and
$c=-0.02$. The other parameters are the same as the figure 1. }
\end{figure}

\vskip 0.3cm

{\noindent \textbf{4. }\textbf{Conclusion}}

 \vskip 0.3cm

In summary, we present a family of single- and two-soliton
solutions of BEC under a periodically oscillating harmonic
potential by using Darboux transformation. It is found that a
single bright soliton exhibits nonperiodic oscillation for a slow
oscillating harmonic potential, while its propagation properties
are hardly affected by the fast oscillation potential.
Furthermore, for two bright solitons, a head-on collision takes
place under a slow oscillating harmonic potential, while there
occurs a ``chase" collision under a fast oscillating harmonic
potential. The collisional behavior can be controlled by the
oscillating frequency of harmonic potential. The results will
stimulate experiments to manipulate solitons in the BEC.
\vskip0.3cm
 \noindent \textbf{Acknowledgments}
 \vskip 0.2cm
This work was supported by National Natural Science Foundation of
China (No. 10674113 and 11074212), Foundation for the Author of
National Excellent Doctoral Dissertation of China (Grant No.
200726), Hunan Provincial Innovation Foundation for Postgraduate
(Grant No. CX2010B254), and the Science and Technology Foundation
of Guizhou Province (Grant No. J20112219).

\end{document}